\documentclass[twocolumn]{aastex63}
\usepackage{amsmath}
\usepackage{color}
\usepackage{graphicx}
\usepackage{subfigure}
\usepackage{float}
\usepackage{textcomp}
\usepackage[T1]{fontenc}
\usepackage{ae,aecompl}
\usepackage{amssymb}
\usepackage{booktabs}
\def\beq{\begin{eqnarray}}
\def\eeq{\end{eqnarray}}

\begin{document}

\title{Probing blackbody components in gamma-ray bursts from black hole neutrino-dominated accretion flows}

\correspondingauthor{Tong Liu}
\email{tongliu@xmu.edu.cn}

\author[0000-0002-6588-2652]{Xiao-Yan Li}
\affiliation{Department of Astronomy, Xiamen University, Xiamen, Fujian 361005, China}

\author[0000-0001-8678-6291]{Tong Liu}
\affiliation{Department of Astronomy, Xiamen University, Xiamen, Fujian 361005, China}

\author[0000-0002-4448-0849]{Bao-Quan Huang}
\affiliation{Department of Astronomy, Xiamen University, Xiamen, Fujian 361005, China}

\author[0000-0002-6727-7798]{Guo-Yu Li}
\affiliation{Laboratory for Relativistic Astrophysics, Department of Physics, Guangxi University, Nanning 530004, China}

\author[0000-0003-1474-293X]{Da-Bin Lin}
\affiliation{Laboratory for Relativistic Astrophysics, Department of Physics, Guangxi University, Nanning 530004, China}

\author[0000-0002-0926-5406]{Zhi-Lin Chen}
\affiliation{Laboratory for Relativistic Astrophysics, Department of Physics, Guangxi University, Nanning 530004, China}

\author[0000-0002-8385-7848]{Yun Wang}
\affiliation{Key Laboratory of Dark Matter and Space Astronomy, Purple Mountain Observatory, Chinese Academy of Sciences, Nanjing 210034, China}

\begin{abstract}
A stellar-mass black hole (BH) surrounded by a neutrino-dominated accretion flow (NDAF) is generally considered to be the central engine of gamma-ray bursts (GRBs). Neutrinos escaping from the disk will annihilate out of the disk to produce the fireball that could power GRBs with blackbody (BB) components. The initial GRB jet power and fireball launch radius are related to the annihilation luminosity and annihilation height of the NDAFs, respectively. In this paper, we collect 7 GRBs with known redshifts and identified BB components to test whether the NDAF model works. We find that, in most cases, the values of the accretion rates and the central BH properties are all in the reasonable range, suggesting that these BB components indeed originate from the neutrino annihilation process.
\end{abstract}


\section{Introduction}

Gamma-ray bursts (GRBs) are the most luminous transients in the Universe. Observationally, most of the GRB prompt emission spectra can be well fitted with the non-thermal Band function, which consists of two empirical smoothly connected power laws \citep[e.g.,][]{Band1993ApJ}. However, prominently thermal or quasi-thermal emission have been observed in several bursts, such as GRB~090902B \citep[e.g.,][]{Abdo2009ApJ,Ryde2010ApJ}, GRB~100724B \citep[e.g.,][]{Guiriec2011ApJ}, GRB~101219B \citep[e.g.,][]{Larsson2015ApJ}, GRB~110721A \citep[e.g.,][]{Axelsson2012ApJ}, GRB 210610B \citep[e.g.,][]{SongXinying2024ApJ}, and so on. Compared to the Band function, the thermal spectrum is much narrower, and it is much harder at low energies.

To interpret the observations of GRB spectra, two types of jets are proposed, i.e., a matter-dominated fireball and a non-thermal Poynting-flux-dominated jet. The former case corresponds to a hot fireball consisting of hot photons, electron/positron pairs and baryons. The optical depth at the initial radius of the fireball $R_0$ is much larger than unity and the thermal photons escape around the fireball photosphere $R_{\rm{ph}}$ where the optical depth $\tau \sim 1$ and thus forms the observed thermal emission \citep[e.g.,][]{Goodman1986ApJ,Paczynski1986ApJ,Shemi1990ApJ}. Besides, a fraction of the thermal energy of the fireball will be converted into the kinetic energy of the jet, which may be dissipated in internal/external shocks and converted into the non-thermal radiation components. On the other hand, a non-thermal component can also be produced by the Poynting-flux-dominated jet.

One of the competitive models of GRB central engines is a rotating black hole (BH) surrounded by a neutrino-dominated accretion flow (NDAF). The physics of NDAFs have been extensively studied in the literature \citep[e.g.,][]{Popham1999ApJ,Kohri2002ApJ,Rosswog2003MNRAS,GuWeimin2006ApJ,Janiuk2007ApJ,Kawanaka2007ApJ,LiuTong2007ApJ,Zalamea2011MNRAS}, for a review see \citet{LiuTong2017NewAR}. Neutrinos can exploit the thermal energy of the heated disk and liberate enormous binding energy, and then neutrinos and anti-neutrinos annihilate in the space out of the disk to produce a hot fireball. Meanwhile, the Poynting jets can be driven by the Blandford-Znajek (BZ) mechanism \citep{Blandford1977MNRAS}. For the BH hyperaccretion, the GRB jet energy is supplied by the BH's rotational energy \citep[e.g.,][]{Lee2000PhR,ZhangBing2011ApJ,LiuTong2015ApJS}. Of course, neutrino annihilation process and BZ mechanism may coexist in the BH hyperaccretion system.

In this paper, we use the GRB blackbody (BB) emission to test the NDAF model under the fireball assumption. This paper is organized as follows. In Section 2, we describe the calculations of the fireball launch radius and the annihilation height. The results are shown in Section 3. Conclusions and discussion are made in Section 4.

\section{Method} \label{sec:model}

\subsection{Fireball launch radius}

Thermal components have been identified in several GRBs. For these bursts, we calculate the fireball launch radius  $R_0$ following the approach in \citet{Pe'er2007ApJ}. The observed BB radiation flux is read as
\begin{eqnarray}
F^{\rm{ob}}_{\rm{BB}} = (2\pi/d_L^2) \int d\mu \mu  R_{\rm{ph}}^2 \mathcal {D}^4  (\sigma T'^4 / \pi).
 \label{eq1}
\end{eqnarray}
The terms on the right-hand side represent the integration of the intensity over the emitting surface. Here, $d_L$ is the luminosity distance, $\mu \equiv \cos\theta$ with $\theta$ being the angle between the direction of the outflow and the line of sight, $\mathcal {D} \approx 1.48\Gamma $ is the Doppler factor with $\Gamma$ being the Lorentz factor, $\sigma$ is Stefan's constant, and $T'$ is the temperature at the photospheric radius $R_{\rm{ph}}$. The quantities with prime sign are in the comoving-frame.

Integrating the Equation~(\ref{eq1}) from $\theta=0$ to $\theta=1/\Gamma$, the effective size of the emission region, defined as $\mathcal{R}= (F_{\rm{BB}}^{\rm{ob}}/\sigma {T^{\rm{ob}}}^4)^{1/2}$, can be derived as
\begin{eqnarray}
\mathcal{R}=1.06 \frac{(1+z)^2}{d_L} \frac{R_{\rm{ph}}}  {\Gamma},
\label{eq2}
\end{eqnarray}
where the factor $1.06$ is due to the geometric effect, $z$ is the redshift, and ${T^{\rm{ob}}}=\mathcal {D}T'/(1+z)$ is the temperature of the thermal component.

For the situation $R_{\rm{ph}}>R_{\rm{s}}$, $R_0$ can be determined. Here, $R_{\rm{s}}$ is the saturation radius of the fireball. In this case, $R_{\rm{ph}}=L \sigma_T /(8\pi \eta^3  m_p c^3)$. In addition, one has $T =T_0 $ and $T \propto r^{-2/3}$ during the acceleration phase ($r < R_{\rm{s}}$) and coasting phase ($R_{\rm{s}}<r<R_{\rm{ph}}$), respectively. Thus,
\begin{eqnarray}
T'_{\rm{ph}} =  T_0 \eta^{-1} \left( \frac{R_{\rm{ph}}}{R_{\rm{s}}}\right)^{-2/3}.
\label{eq3}
\end{eqnarray}
Combining with $T_0=[L/4 (\pi R_0^2 ca)]^{1/4} $ being the BB temperature at $R_0$ and $L = 4 \pi d_L^2 Y F^{\rm{ob}}$, one can finally obtain
\begin{eqnarray}
R_0 = 0.6\frac{d_L} {(1+z)^2} \left( \frac{F^{\rm{ob}}_{\rm{BB}} }{ Y F^{\rm{ob}}} \right)^{3/2} \mathcal{R},
\label{eq4}
\end{eqnarray}
where $F^{\rm{ob}}$ is the total observed energy flux in $\gamma$-rays (both thermal and non-thermal), and  $1/Y$ represents the $\gamma$-ray production efficiency of the fireball, i.e.,
\begin{eqnarray}
Y =  \frac{E_{\gamma, \mathrm{iso}}+E_{\mathrm{k}, \mathrm{iso}}}{E_{\gamma, \mathrm{iso}}},
\label{eq5}
\end{eqnarray}
where $E_{\gamma, \mathrm{iso}}$ and $E_{\mathrm{k}, \mathrm{iso}}$ are the isotropic radiated energy in the prompt emission and the isotropic kinetic energy to power GRB afterglows, respectively.

Note that $Y=1$ is usually assumed in the literature because of the large uncertainties in estimating $E_{\mathrm{k}, \mathrm{iso}}$. However, we find that the values of $Y$ significantly affect the results of $R_0$ by about an order of magnitude. Thus, in our work, $R_0$ is recalculated by considering the value of $Y$ for each burst.

\subsection{Annihilation height}

\citet{XueLi2013ApJS} calculated one-dimensional global solutions of NDAFs by considering general relativity, neutrino physics, and nucleosynthesis in detail. They find that in the range of $0.01 < \dot{m} < 0.5$ (the range can be extended to $\dot{m} \lesssim 1$), the annihilation luminosity and annihilation height can be approximately written as
\begin{eqnarray}
\begin{aligned}
\log L_{\nu \bar{\nu}}~({\rm erg~s^{-1}})=& 52.98 + 3.88  a_* - 1.55  \log m_{\rm BH} \\+& 5.0 \log \dot{m},
\label{eq6}
\end{aligned}
\end{eqnarray}
and
\begin{eqnarray}
\log h  =  2.15 - 0.30 a_* - 0.53 \log m_{\rm BH} +  0.35 \log \dot{m},
\label{eq7}
\end{eqnarray}
where $m_{\rm BH}=M_{\rm BH}/M_\odot$ and $\dot{m}=\dot{M}/M_\odot~\rm s^{-1}$ are respectively the dimensionless BH mass and accretion rate, $0\leq a_* \leq 1$ is the dimensionless BH spin parameter, $h=H/r_{\rm g}$, and $r_{\rm g}=2GM_{\rm BH}/c^2$ is the Schwarzschild radius. Here, the annihilation height is defined as the region where $99.9\%$ of the annihilation luminosity is included.

From the observational point of view, the mean jet power of GRBs can be estimated as \citep[e.g.,][]{FanYizhong2011ApJ,LiuTong2015ApJ}
\begin{eqnarray}
P_{\mathrm{j}} \simeq \frac{\left(E_{\gamma, \mathrm{iso}}+E_{\mathrm{k}, \mathrm{iso}}\right)(1+z) \theta_{\mathrm{j}}^{2}}{2 T_{90}},
\label{eq8}
\end{eqnarray}
where $\theta_{\mathrm{j}}$ is the half-opening angle of the jet and $T_{90}$ is the burst duration. Using these observational GRB data, the jet power $P_{\mathrm{j}}$ can be obtained.

Here we adopt $P_{\rm j}=L_{\nu\bar\nu}$. Based on Equations~(\ref{eq6}) and (\ref{eq7}), for typical ranges of the BH mass, i.e., $3 \leq m_{\rm BH}\leq 10$, one can obtain the annihilation height to contrast the fireball launch radius.

\section{results} \label{sec:results}

\begin{table*}
\centering
\caption{Collected GRB Data}
\resizebox{2.2\columnwidth}{!}{
\begin{tabular}{cccccccccccccc}
\hline
\hline
GRB & $z$ & $T_{90}$  & $E_{\rm \gamma,\rm iso}$ & $E_{\rm k,\rm iso}$&$\theta_{\rm j}$ &$kT$ & $F_{\rm{BB}}$ & $F_{\rm{BB}}/F_{\rm {tot}}$ & $Y$ & $R_0$ & $P_{\rm j}$ & Ref&\\
~ &~ & (s) & ($10^{52}\;\rm erg$) & ($10^{52}\;\rm erg$) &(rad)&$(\rm{keV})$ &$(\rm {erg\;cm^{-1}}\;s^{-1})$& & &($10^{7}\;\rm \rm{cm}$)& ($10^{49}\; \rm erg\rm~s^{-1}$)   \\ \hline
970828 & 0.96 & 160 & 29 & 37.154 &0.12&$78.5 \pm {4}$ &$1.38 \times 10^{-6} $&0.64& 2.3 & 8.88 & 6.22 & 1,2 \\
990510 & 1.619 & 67.58 & 17 & 13.16 &0.06&$46.5 \pm {2}$& $7.0 \times 10^{-7} $&0.25 & 1.8 & 6.09& 2.01 & 1,2 \\
080810 & 3.355 & 108 & 30 & 1.6&\textgreater0.07&62&$1.6 \times 10^{-7}$& 0.28& 1.1 & 2.33 & 3.1 & 1,3 \\
090902B & 1.8829 & 19.328 & 320& 56&0.068 &168&$1.96 \times 10^{-5}$&0.26& 1.2 &5.19 & 130 & 1,4 \\
101219B & 0.552 & 51 & 0.34 &6.4&\textgreater0.3&$19.1\pm 0.7$&$8.45\times 10^{-8}$&$ \lesssim 1$& 20.0 & 2.36 & 9.11 & 1,2 \\
190114C & 0.425 & 116.354 & 41.2& 190&\textgreater 0.57&159&$2.25\times 10^{-5}$&0.2&  5.6 &0.29& 455 & 2,5 \\
110731A*& 2.83 & 7.3 & 68 & 7.5 &\textgreater0.15&$85\pm 5$&$6.54 \times 10^{-6}$&0.47& 1.1 & 30.0 & 472.06 & 1,2 \\
\hline
\end{tabular}}
\begin{minipage}{16cm}
\footnotesize
\emph{References}: \\
(1) \citet{Pe'er2015ApJ};
(2) \citet{LiXiaoyan2024MNRAS};
(3) \citet{Page2009MNRAS};
(4) \citet{Cenko2011ApJ};
(5) \citet{Moradi2021PhRvD}.\\
\end{minipage}
\label{MyTabA}
\end{table*}

Assuming that these bursts are matter-dominated, we collect 7 GRBs with known redshift and identified thermal components to calculate the fireball launch radius and the annihilation height. Besides, the bursts with inconclusive or controversial BB components are not included. For example, \citet{Izzo2012AA} fitted the spectral of GRB~090618 and found that the BB with a power law model can fit well for the first 50 $\rm{s}$ of emission. However, the emission can also be well-fitted by a Band model. In addition, GRB~210610B may have an apparent BB component \citep[e.g.][]{SongXinying2024ApJ}, but $Y$ cannot be obtained due to the lack of fitting of the afterglows. For GRB~100724B, the redshift is unknown. All of these bursts are not considered in this paper. The redshift $z$, duration $T_{90}$, opening angle $\theta_{\rm j}$, isotropic radiated energy in the prompt emission phase $E_{\gamma, \mathrm{iso}}$, isotropic kinetic energy $E_{\mathrm{k}, \mathrm{iso}}$, thermal temperature $kT$, thermal flux $F_{\rm{BB}}$ and the ratio of thermal flux to the total flux $F_{\rm{BB}}/F_{\rm{tot}}$ are collected in Table~\ref{MyTabA}. The jet launch radius $R_0$, the energy ratio $Y$, and the jet power $P_{\rm{j}}$ derived from Equations~(\ref{eq4}), (\ref{eq5}), and (\ref{eq8}) are also presented. It should be noticed that the measurements of $E_{\rm \gamma,\rm iso}$ and  $E_{\rm k,\rm iso}$ are model dependent. Besides, the lower limits of $\theta_{\rm j}$ are used in some bursts. Another uncertainty comes from the fitting of thermal spectrum. Thus, one should notice that the values of $R_0$ are only approximate results.

The work is conducted in the following procedure. First, we set $m_{\rm BH}\in[3, 10]$ while $\dot{m}$ is set to be free. Then, $a_*$ is  calculated with  Equation (\ref{eq6}), and only the values $a_*\in[0, 1]$ are reserved. Finally, $H$ can be obtained with Equation~(\ref{eq7}) with the known jet luminosity. In Figure \ref{MyFig}, we illustrate the parameter space for GRB samples. The multicolor regions denote the range of the dimensionless accretion rates and $\dot{m}=0.2$ are marked with dash-dotted lines. The black solid lines represent $a_*$ = 0.1, 0.5, and 0.9. Besides, the launch radii $R_0$ of the bursts are plotted with black dashed lines. One should notice that $H \lesssim R_0$ will be satisfied. The neutrinos emitted from the disk will annihilate with anti-neutrinos outside the disk to produce electron-positron pairs which later annihilate into photons. Some electron-positron pairs can escape if they beyond the ``stagnation surface'' above which they will not be swallowed by the BH. These pairs will further be accelerated and finally generate a collimated jet \citep[e.g.][]{McKinney2005astro}. Thus, the neutrino annihilation height should be less than the initial radius of the fireball.

\begin{itemize}
\item
GRB~970828, located at R.A.(J2000) = $18^{\rm{h}} 08^{\rm{m}} 29^{\rm{s}}$, and decl.(J2000) = $+59^\circ 18'0''$, triggered the ASM aboard \emph{RXTE} on 28 August 1997 at  $17^{\rm{h}}44^{\rm{m}}36^{\rm{s}}$ UT \citep[e.g.,][]{Groot1998ApJ}. The duration is about 160 $\rm s$ in the 2-12 keV energy band and the redshift $z$ is about 0.96 \citep[e.g.,][]{Djorgovski2001ApJ}. \citet{Pe'er2007ApJ} used a Planck function and a single power law to fit the thermal component and non-thermal component of the time-resolved spectra of the burst, respectively. They find that the temperature rises slightly at first 8 $\rm s $ at $78.5 $ keV then shows a decay phase later. The ratio of the observed thermal flux to the total flux $F_{\rm{BB}}/F_{\rm {tot}}=0.64 \pm 0.2$, and $\mathcal{R} \sim 1.88 \times 10^{-19}$. $E_{\rm \gamma,\rm iso}$ and $E_{\rm k,\rm iso}$ are $2.9 \times 10^{53}\;\rm{erg}$ and $3.7 \times 10^{53}\;\rm{erg}$, respectively \citep[e.g.,][]{LiXiaoyan2024MNRAS}. With these values, we obtain the initial size of the flow $R_0=8.88 \times 10^{7}\;\rm {cm}$, which is almost covered by the reasonable region of $H$.

\item
GRB~990510, located at R.A.= $13^{\rm{h}} 38^{\rm{m}} 06^{\rm{s}}$, and decl.(J2000) = $-80^\circ 29'30''$, triggered the \emph{BeppoSAX} on 10 May 1999 at  $10^{\rm{h}} 22^{\rm{m}} 02^{\rm{s}}$ UT. The duration is approximately 67.58 $\rm s$ and the redshift $z$ is about 1.619 \citep[e.g.,][]{Vreeswijk2001ApJ,Pe'er2007ApJ}. \citet{Pe'er2007ApJ} conducted a similar analysis as GRB~970828 for this burst and they found that the BB temperature is 46.5  keV. The thermal flux $F_{\rm{BB}}=7 \times 10^{-7} \;\rm {erg\;cm^{-1}\;s^{-1}}$ and the ratio of the observed thermal flux to the total flux $F_{\rm{BB}}/F_{\rm{tot}}=0.25$. $E_{\rm \gamma,\rm iso}$ and $E_{\rm k,\rm iso}$ are $1.7 \times 10^{53}\;\rm{erg}$ and $1.3 \times 10^{53}\;\rm{erg}$, respectively \citep[e.g.,][]{LiXiaoyan2024MNRAS}. Thus, one can obtain the initial size of the flow $R_0=6.09 \times 10^{7} \;\rm{cm} $, which is also in the reasonable region of $H$.

\item
GRB~080810, located at R.A.(J2000) = $23^{\rm{h}} 47^{\rm{m}} 10^{\rm{s}}.48$, and decl.(J2000) = $+00^\circ 19'11.3''$, is a long GRB at redshift $z = 3.355$ triggered  the \emph{Swift} and \emph{Fermi} on 10 August 2008 at  $13^{\rm{h}} 10^{\rm{m}} 12^{\rm{s}}$ UT. The duration $T_{90}$ is 108 $\rm s$ \citep[e.g.,][]{Page2009MNRAS}. They fitted the time-sliced spectra and found that the BB temperature is $62$ keV. The thermal flux $F_{\rm{BB}}=1.6 \times 10^{-7}\;\rm{erg\;cm^{-1}\;s^{-1}}$ and the ratio of the observed thermal flux to the total flux $F_{\rm{BB}}/F_{\rm{tot}}=0.28$. $E_{\rm \gamma,\rm iso}$ and $E_{\rm k,\rm iso}$ are $3 \times 10^{53} \;\rm{erg}$ and $1.6 \times 10^{52} \;\rm{erg}$, respectively. Thus, one can obtain the initial size of the flow $R_0=2.33 \times 10^{7}\;\rm{cm}$, which is completely in the range of $H$.

\begin{figure*}
\centering
\includegraphics[width=0.45\textwidth,height=0.3\textwidth]{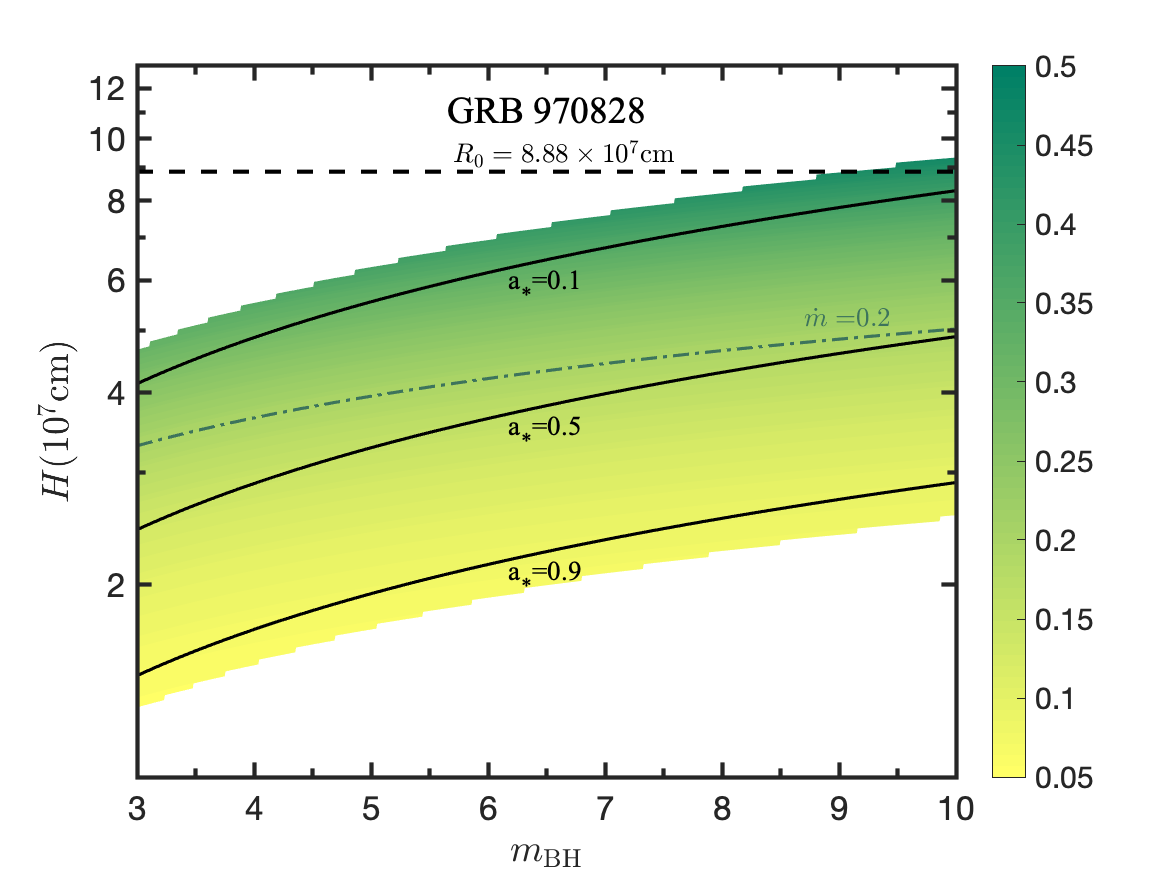}
\includegraphics[width=0.45\textwidth,height=0.3\textwidth]{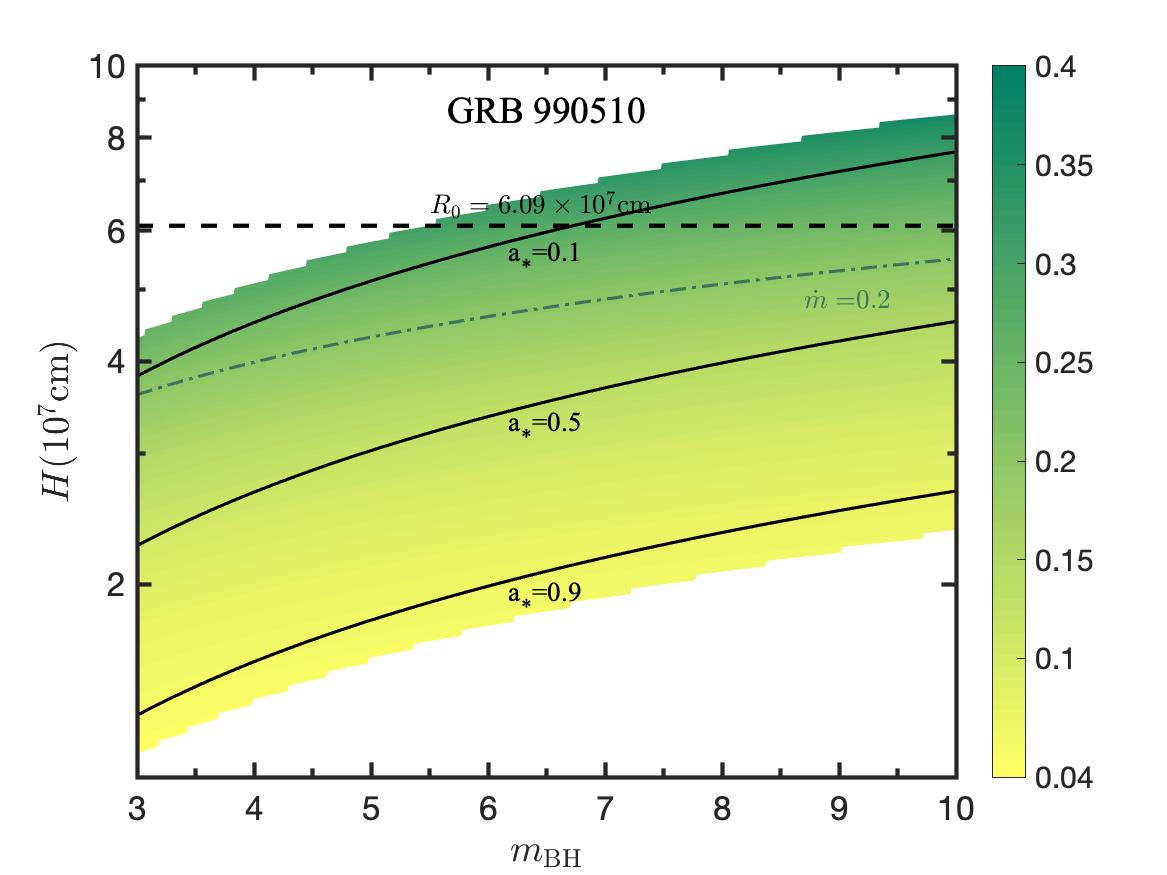}
\includegraphics[width=0.45\textwidth,height=0.3\textwidth]{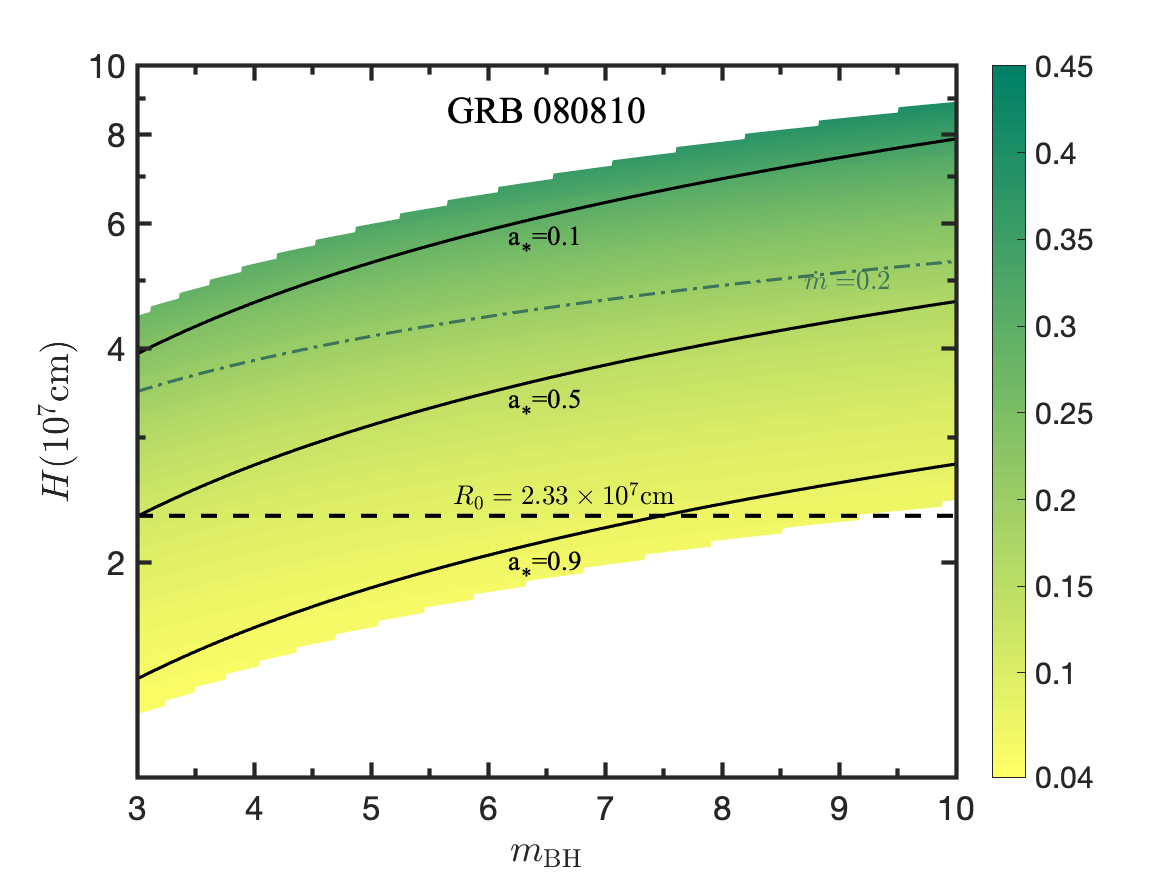}
\includegraphics[width=0.45\textwidth,height=0.3\textwidth]{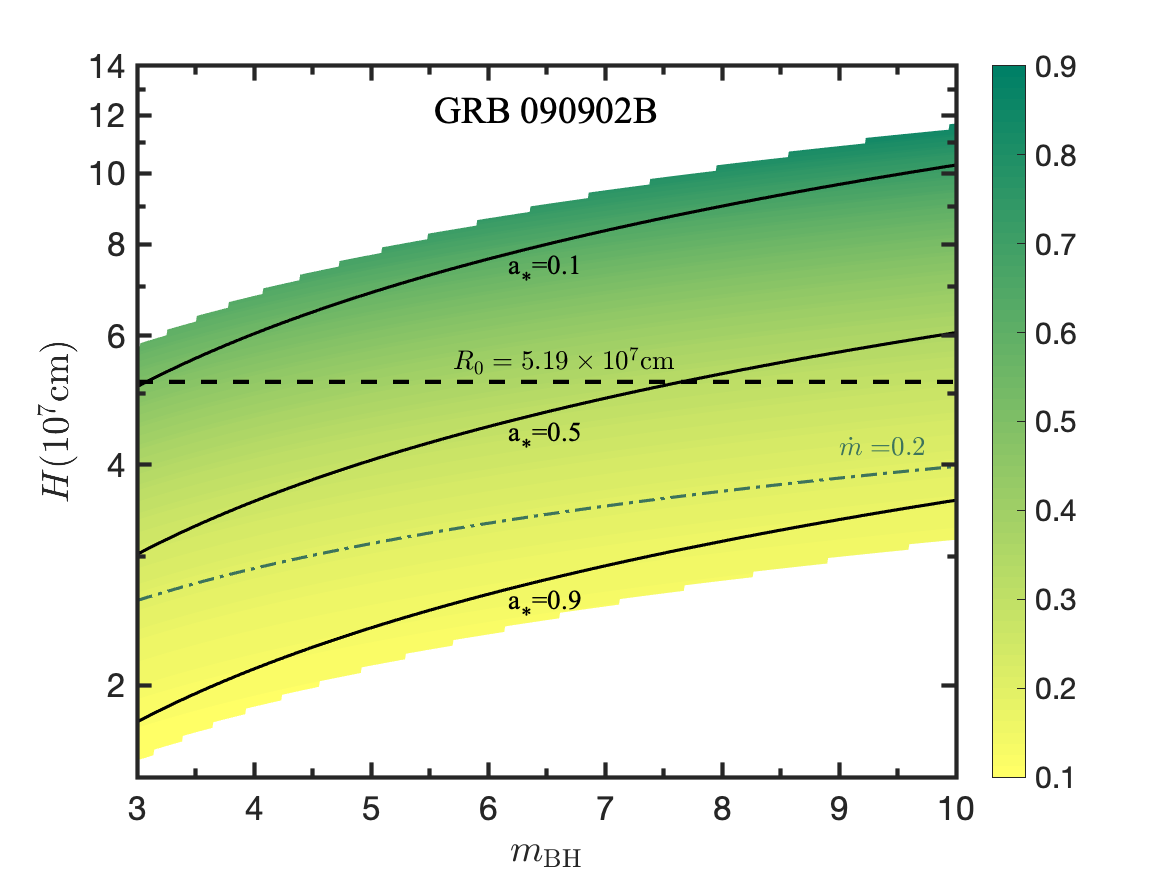}
\includegraphics[width=0.45\textwidth,height=0.3\textwidth]{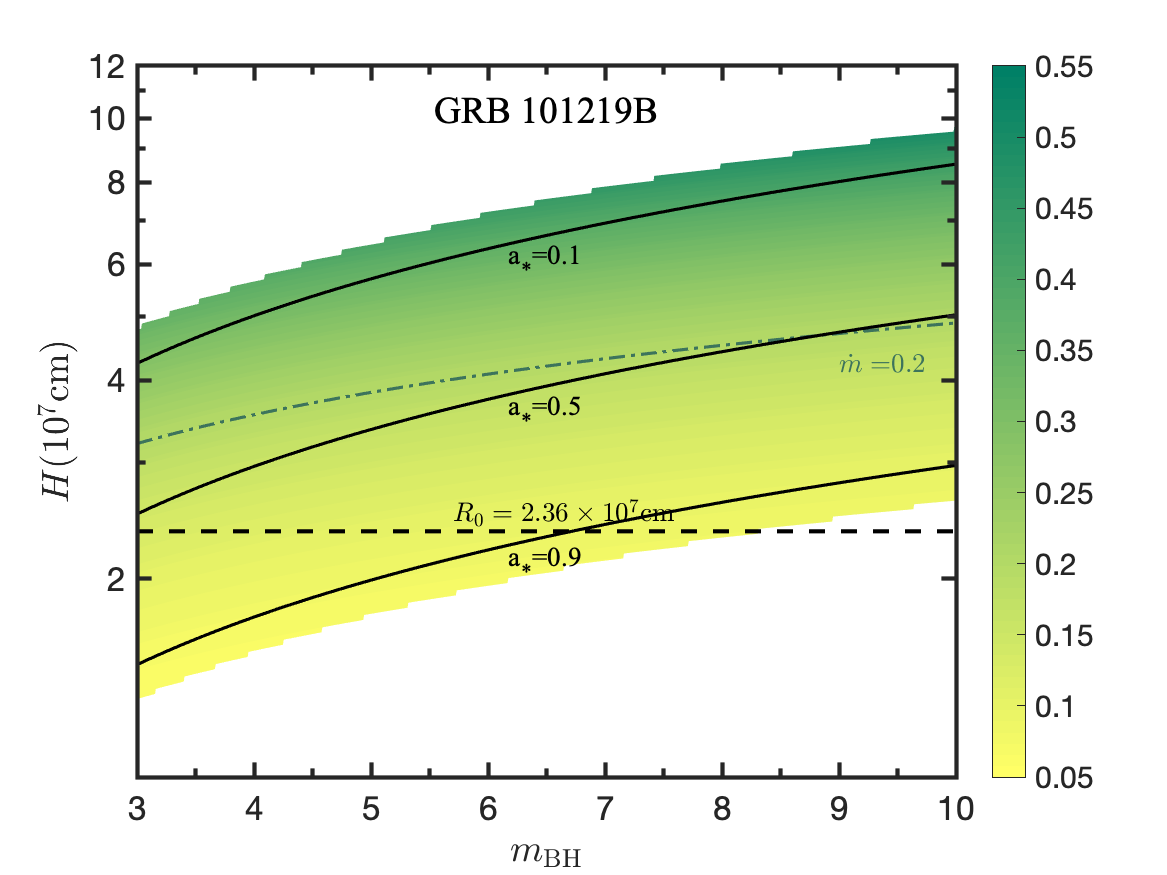}
\includegraphics[width=0.45\textwidth,height=0.3\textwidth]{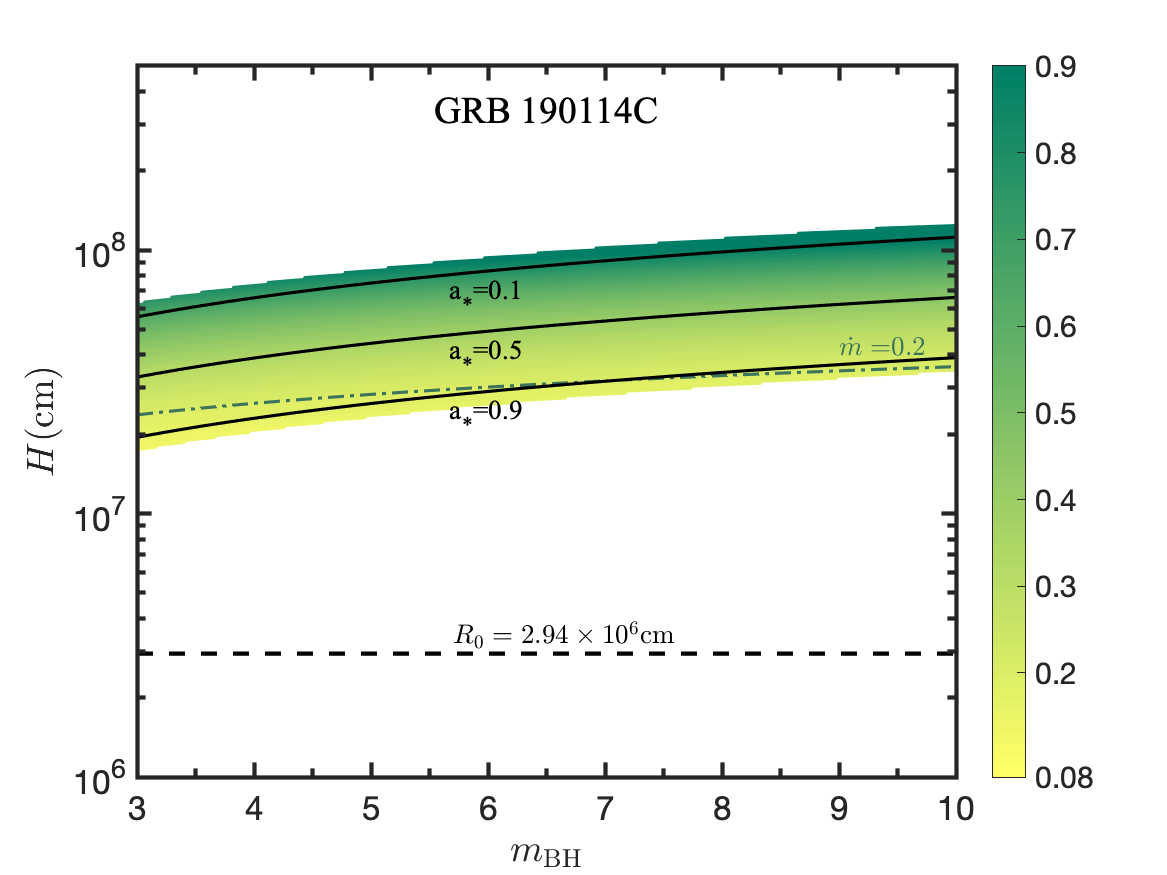}
\includegraphics[width=0.45\textwidth,height=0.3\textwidth]{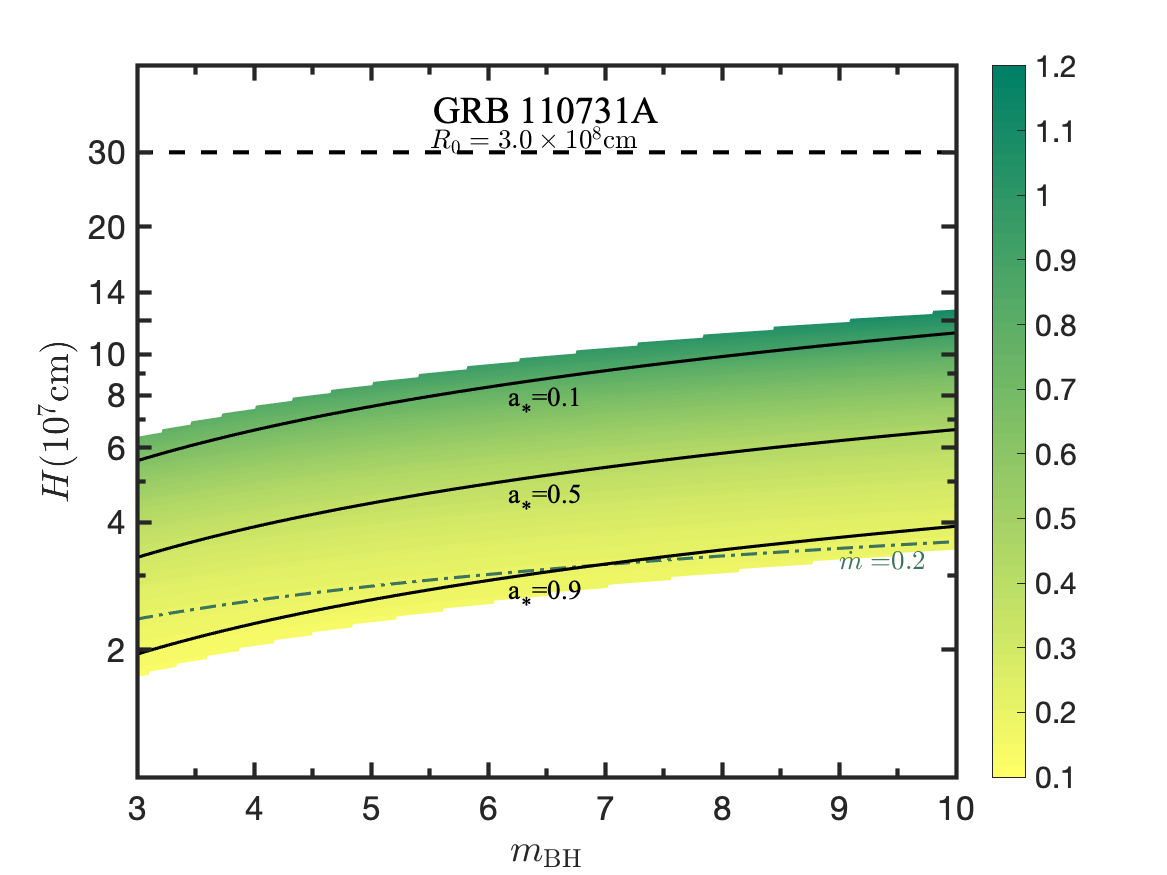}
\caption{Neutrino annihilation height $H$ as a function of BH properties and accretion rates. The multicolor regions denote the range of the dimensionless accretion rates and the green dash-dotted lines correspond to $\dot{m}=0.2$. The black solid lines correspond to the different BH spin. The black dashed lines correspond to the launch radii of the GRB fireballs.}
\label{MyFig}
\end{figure*}

\item
GRB~090902B is one of the brightest long bursts detected by \emph{Fermi}/LAT with redshift $z = 1.8829$ and $T_{90}=19.328 \;\rm s$ located at R.A.(J2000) = $17^{\rm{h}} 40^{\rm{m}} 00^{\rm{s}}$ and decl.(J2000) = $+27^\circ 19'48''$ \citep[e.g.,][]{Abdo2009ApJ}. GRB~090902B is a thermal-dominated burst since a narrow spectrum has been observed. The BB temperature is $168$ keV. The thermal flux $F_{\rm{BB}}=1.96 \times 10^{-5} \;\rm {erg\;cm^{-1}\;s^{-1}}$ and the ratio of the observed thermal flux to the total flux $F_{\rm{BB}}/F_{\rm{tot}} =0.26$ \citep{Pe'er2015ApJ}. $E_{\rm \gamma,\rm iso}$ and $E_{\rm k,\rm iso}$ are $3.2 \times 10^{54}\;\rm{erg}$ and $5.6 \times 10^{53}\;\rm{erg}$, respectively \citep[e.g.,][]{Cenko2011ApJ}. Thus, one can obtain the initial size of the flow $R_0=5.19 \times 10^{7}\;\rm{cm}$, which is  within the reasonable range of $H$.

\item
GRB 101219B, located at R.A.(J2000) = $00^{\rm{h}} 49^{\rm{m}} 02^{\rm{s}}$, and decl.(J2000) = $-34^\circ 31' 53''$, is a long GRB triggered both \emph{Swift}/BAT  and \emph{Fermi}/GBM at $16^{\rm{h}} 27^{\rm{m}} 53^{\rm{s}}$ UT. The duration $T_{90}$ is 51 $\rm s$, and the redshift $z$ is 0.552 \citep[e.g.,][]{Larsson2015ApJ}. The BB temperature is $19.1$ keV. The thermal flux $F_{\rm{BB}}=8.45\times 10^{-8} \;\rm {erg\;cm^{-1}\;s^{-1}}$ and the ratio of the observed thermal flux to the total flux $F_{\rm{BB}}/F_{\rm{tot}}\lesssim 1$. $E_{\rm \gamma,\rm iso}$ and $E_{\rm k,\rm iso}$ are $3.4 \times 10^{51} \;\rm{erg}$ and $6.4 \times 10^{52} \;\rm{erg}$, respectively \citep[e.g.,][]{LiXiaoyan2024MNRAS}. Thus, one can obtain the initial size of the flow $R_0=2.36 \times 10^{7} \;\rm {cm}$, which requires a low accretion rate, a high BH spin, and a $\lesssim 9~M_\odot$ BH \citep{LiuTong2016ApJ}.

\item
GRB~190114C triggered \emph{Fermi} on 14 January 2019 at $20^{\rm{h}} 57^{\rm{m}} 03^{\rm{s}}$, UT. The location of the burst is R.A. =  $03^{\rm{h}} 38^{\rm{m}} 02^{\rm{s}}$, and decl. =$-26^\circ 56' 18''$ \citep[e.g.,][]{Gropp2019GCN,Hamburg2019GCN}. The duration $T_{90}$ is about 116 $\rm s$  and the redshift $z = 0.425$. The BB temperature is $159$ keV. The thermal flux $F_{\rm{BB}}=2.25 \times 10^{-5} \;\rm {erg\;cm^{-1}\;s^{-1}}$ and the ratio of the observed thermal flux to the total flux $F_{\rm{BB}}/F_{\rm{tot}}=0.2$ \citep{Moradi2021PhRvD}. $E_{\rm \gamma,\rm iso}$ and $E_{\rm k,\rm iso}$ are $4.12 \times 10^{53} \;\rm{erg}$ and $1.9 \times 10^{54} \;\rm{erg}$, respectively \citep[e.g.,][]{LiXiaoyan2024MNRAS}. Thus, one can obtain the initial size of the flow $R_0=2.94 \times 10^{6}\;\rm{cm}$, which is lower than $H$ in the whole range. Actually, this fireball launch radius is so small that it approaches a $\sim 9~M_\odot$ BH's horizon, the annihilation could not be sufficient in this narrow space out of the disk even for lower-mass BHs and for any accretion rate. Besides NDAF models, it is difficult to explain the fireball launch radius of GRB~190114C by other BB models, such as a cocoon or a jet propagating in a dense environment, which occurs far from the center. Since the launch radius $R_0$ of this burst seems to be much smaller than others, the uncertainties in the estimation of the jet parameters, such as $\theta_{\mathrm{j}}$, $E_{\mathrm{k,iso}}$, and $E_{\gamma, \mathrm{iso}}$, might be the main causes. In addition, the fitting of the BB component may also be inaccurate due to the limitation of the observations. In any case, the value of $R_0$ in GRB 190114C is unconvincing.

\item
GRB~110731A triggered both \emph{Swift}/BAT and \emph{Fermi}/GBM on 31 July 2011 at  $11^{\rm{h}} 09^{\rm{m}} 30^{\rm{s}}$ UT. The location is at R.A.(J2000) = $18^{\rm{h}} 41^{\rm{m}} 00^{\rm{s}}$, and decl.(J2000) = $-28^\circ 31'00''$. The redshift is $2.83$ and the duration $T_{90}$ in the 50-300 keV energy band is 7.3 $\rm s$ \citep{Ackermann2013ApJ}. The BB temperature is about $85$ keV. The thermal flux $F_{\rm{BB}}=6.54 \times 10^{-6} \; \rm {erg\;cm^{-1}\;s^{-1}}$ and the ratio of the observed thermal flux to the total flux $F_{\rm{BB}}/F_{\rm{tot}}=0.47$. $E_{\rm \gamma,\rm iso}$ and $E_{\rm k,\rm iso}$ are $6.8 \times 10^{53} \;\rm{erg}$ and $7.5 \times 10^{52} \;\rm{erg}$, respectively \citep[e.g.,][]{LiXiaoyan2024MNRAS}. Thus, one can obtain the initial size of the flow $R_0=3.0 \times 10^{8}\; \rm{cm}$, which is much higher than $H$ for all of the reasonable parameters. GRB~110731A is a peculiar burst since the duration $T_{90}>2 \;\rm{s}$ but may originate from a compact-star merger \citep{LvHoujun2017ApJ}. The large $R_0$ could be explained by the NDAF model, and could also be related to a cocoon or a jet propagating in a dense environment since they are far from the center.
\end{itemize}

In this work, we assumed the fireball model for these seven bursts. \citet{2015ApJ...801..103G} proposed a ``top-down'' approach to derive the magnetization parameter $\sigma_0=L_{\rm {c,0}}/L_{\rm {h,0}}$, where $L_{\rm {c,0}}$ and $L_{\rm {h,0}}$ are the luminosities of the hot and cold components, respectively. It is an effective method to diagnose the jet composition. Some studies also used this method and found that most GRBs that have no detectable thermal component are consistent with having a strongly magnetized engine \citep[e.g.,][]{2009ApJ...700L..65Z,2020ApJ...894..100L,2024ApJ...972....1L}. Here, for the samples with thermal component detected, we calculate $\sigma_{0}$ by using the Equation (8) of \citet[][]{2024ApJ...972....1L}. Using the $R_{0}$ values obtained above, we estimate that $\sigma_0=0.1$ for GRB~990510, $\sigma_0=0.85$ for GRB~080810, $\sigma_0=0.66$ for GRB~090902B, $\sigma_0=0.95$ for GRB~101219B, $\sigma_0=0.76$ for GRB~110731A, and $\sigma_0$ much less than unity for GRB~970828 and GRB~190114C. The magnetization parameters of all the bursts are less than unity, which indicates that the fireball assumption applied in this work model is reasonable.

For most of the samples above, $R_0 \gtrsim H$ is satisfied with a reasonable range of BH characteristics and accretion rate, which suggests that the neutrino annihilation mechanism is a good explanation for the thermal emission of GRBs \cite[e.g.,][]{LiuTong2015ApJ,LiuTong2016ApJ}. Thus, the appropriateness of the existence of NDAFs can be verified.

\section{Conclusions and Discussion}

Combined with the observational data of GRBs with identified thermal components and known redshift, we investigate the origin of the thermal component by NDAFs. In the toy model, we find that $R_0 \gtrsim H$ for most of the samples, suggesting that the neutrino annihilation mechanism is a good explanation for GRB thermal emission.

In fact, the thermal components may widely exist in GRBs. The non-detection of the thermal component maybe because it is too weak and is masked by the non-thermal emission. Even for significant BB components, inadequate observations will lead to the imprecise fitting of the spectra. For the bursts with a weak BB component, $R_0$ is difficult to constrain and the investigation of the NDAF can not be well conducted. Therefore, only the bursts with identified thermal components are included in this work.

It should be noticed that there are some uncertainties in the calculations. First, the jet half-opening angle $\theta_{\mathrm{j}}$ is usually estimated by the jet break time in X-ray afterglows and it is hard to constrain due to the lack of jet break observations. The lower limits of $\theta_{\mathrm{j}}$ are widely used. Second, $E_{\mathrm{k,iso}}$ and $E_{\gamma, \mathrm{iso}}$ are model dependent. Another uncertainty is from the fitting results of the thermal emission spectra, which will result in the deviation of the $R_0$ values. Third, $T_{90}$ is widely used to represent the activity timescale of GRBs, but these two timescales can be very different.

There are some uncertainties in physics. First, the high accretion rate of BH-NDAF systems should trigger the violent evolution of the BH mass and spin, which further influences the neutrino annihilation luminosity \citep[e.g.,][]{SongCuiying2015ApJ,QuHuimin2022ApJ}. Second, for NDAFs, the disk outflows should be taken into account \citep[e.g.,][]{LiuTong2018ApJ}, which will reduce the accretion rate in the inner region of the disk. Third, as mentioned above, in the BH hyperaccretion system, the neutrino annihilation process could coexist with the BZ mechanism, which might partly or even dominantly contribute to the non-thermal emission. In this case, the accretion rate can be drastically reduced and $Y$ should be fixed \citep[e.g.,][]{LiuTong2015ApJS}. One should also notice that the effective $R_0$ of a fireball can be revised if considering the effect of the reborn of a fireball when the jet penetrates through the star \citep[e.g.,][]{{2007MNRAS.382L..72G}}. For this scenario, $R_0$ may reach to $\sim 10^{11} \rm{cm}$, much larger than $\sim 10^{7} \rm{cm}$ and the values calculated in this paper for a naked engine. The NDAF model is valid for $R_0 \gtrsim H$. Thus, it remains valid for this scenario. Nevertheless, the verification of NDAFs is still extremely important for understanding stellar evolution and explosions.

The thermal emission may also exist in light curves of X-ray afterglows \citep[e.g.,][]{Valan2021MNRAS}, such as GRBs~131030A and 151027A. For the thermal emission detected in X-ray afterglows, the thermal component may also be explained by the neutrino annihilation for the restart of the central engine or outer cocoons. A cocoon can be formed when the jet breaks out of the envelope of the progenitor star \citep[e.g.,][]{Ramirez2002MNRAS,Nakar2017ApJ,LiuTong2018ApJ,Valan2021MNRAS}. The emission from the cocoon surrounding the jet is used to explain the thermal components \citep[e.g.,][]{Ghisellini2007MNRAS,Piro2014ApJ}. Moreover, the thermal component may originate from a dense environment. For example, for a burst associated with a supernova, the jet will interact with the supernova ejecta. Thus, the ejecta will be accelerated and heated and finally produce the thermal emission \citep[e.g.,][]{Ruffini2017AA,Valan2021MNRAS}.

Finally, the BH-NDAF systems are the important sources that can produce MeV neutrino and gravitational wave (GW) radiation \citep[e.g.,][]{Suwa2009PhRvD,SunMouyuan2012ApJ,Liu2016}. It is gratifying that with the launch of the Einstein Probe \citep[EP,][]{Yuan2022} and the Space Variable Objects Monitor \citep[SVOM,][]{Bernardini2021,Attiea2022}, the joint observations on GRBs with BB components and associated kilonovae, supernovae, GWs, and neutrino radiation become more possible in the era of multi-messenger astronomy. In the future, the existence of NDAFs will be further verified.

\acknowledgments
We thank the anonymous referee for very helpful suggestions, and Yun-Feng Wei, Sen-Yu Qi, Guo-Peng Li, and Qi Wang for useful discussion. This work was supported by the National Key R\&D Program of China (Grant No. 2023YFA1607902), the National Natural Science Foundation of China (Grant Nos. 12173031, 12221003, and 12273005),  the China Postdoctoral Science Foundation (Grant No. 2024M751769), and the Postdoctoral Fellowship Program of CPSF (Grant No. GZC20241916).

\end{document}